\documentstyle{article}

\begin{document}

\title{Introduction to light forces, atom cooling, and atom trapping.}
\author{Craig Savage \\ Dept. of Physics, Faculty of Science,
ANU, ACT 0200 \\ Craig.Savage@anu.edu.au \\ arch-ive/9510004}

\maketitle

\begin{abstract}
This paper introduces and reviews light forces, atom cooling and atom
trapping.  The emphasis is on the physics of the basic processes.  In
discussing conservative forces the semi-classical dressed states are
used rather than the usual quantized field dressed states.
\end{abstract}

\section{Forces} \label{sec1}
The idea that light exerts mechanical forces on matter arose early on
in astronomy.  In 1619 Kepler suggested that it was the pressure of
sunlight that made comet tails stream away from the Sun (Minogin and
Letokhov 1987).  And he was right.  Nearly 250 years later Maxwell's
theory of electromagnetism quantified light pressure.  The light force
on individual atoms was observed by Frisch, of Meitner and Frisch
fission fame, in 1933.  On Earth lasers provide the intensities
necessary for exerting useful light forces.  In the early 1970s Arthur
Ashkin at Bell Labs accelerated, levitated and trapped micron sized
plastic spheres (Ashkin 1972).  This work has developed into optical
tweezers for manipulating small objects.

Today atoms are routinely cooled and trapped.   Light based mirrors,
beamsplitters, and lenses for atoms have been demonstrated (Adams et
al.  1994a).  These are the subject matter of atom optics in which
atomic de Broglie waves are manipulated as in classical optics.
Atomic interferometers constructed with atom optics have extraordinary
sensitivity (Adams {\em et al.}  1994b).  Promising applications include
atomic clocks, lithography, optical tweezers, and atom lasers or
bosers (Wiseman and Collett 1995).

In this section we introduce the absorptive and dispersive
limits of light forces.  Absorptive forces are dissipative, and hence
useful for cooling, while purely dispersive forces are conservative
and hence useful for maintaining coherence.  Section \ref{sec2} deals
with cooling: the first step towards trapping.  Cooling has two
important milestones: the doppler limit and the recoil limit.  At each
``limit'' a new approach to cooling must be adopted.  At present there
appears to be no fundamental lower limit to the temperatures which can
be achieved.  Section \ref{sec3} surveys methods for trapping ions and
atoms.  For neutral atoms we survey three basic approaches:
dissipative traps, conservative traps, and magnetic traps.

\subsection{Classical electrodynamics}
The electric field of an electromagnetic wave sets a charged
particle oscillating.  The Lorentz interaction with the wave's
magnetic field then pushes the charge in the direction of wave
propagation.  Since the wave gives momentum to the charge it must
itself have momentum.  Relativistically, the presence of energy and
momentum in the energy-momentum 4-vector means that field energy in
some frame is momentum in another frame.

A recent experiment at the University of Queensland demonstrated the
transfer of field angular momentum to micron sized particles
(He {\em et al.} 1995).  The experiment was interesting because it was
electromagnetic orbital angular momentum rather than photon
spin that was transferred (Allen {\em et al.} 1992).

The field's intensity $I$ (W/m$^{2}$) is given by the modulus of the
Poynting vector $\underline{P} = c^{2} \epsilon_{0}\underline{E}
\times \underline{B}$.  Using $p=E/c^{2}$ to convert this to a
momentum flux gives the radiation pressure (N/m$^{2}$):
\begin{equation}
	\underline{R}= c \epsilon_{0}\underline{E} \times \underline{B}, \quad
	| \underline{R} | = I/c.
	\label{fieldmom}
\end{equation}
The major macroscopic application of this force may well turn out to
be solar sailing (Mallove and Matloff 1989).
The solar flux at the Earth's orbit is about 1400 W/m$^{2}$,
corresponding to a pressure of 4.7 $\mu$N/m$^{2}$.  Perfect reflection
doubles this pressure.  Aluminium coated mylar films a few microns
thick have densities of around 5 g/m$^{2}$ giving a payload free
acceleration of around 2 mm/s$^{2}$.  This is enough for
interplanetary travel, taking about a year to reach Mars.

The most useful macroscopic light force today is the dispersive force
used in optical tweezers.  An object with higher refractive
index than its surroundings is attracted towards high field regions,
while one with lower index is repelled.  This can be understood by
considering the momentum changes of light rays refracted through the
particles, Fig.  \ref{fig1}.
In section \ref{conservative} we shall derive the corresponding results for
atoms.  A field detuned above resonance, which induces a refractive
index $n<1$, repels atoms from high field regions, while a below
resonance field, which induces a refractive index $n>1$,
attracts atoms (Loudon 1973).

Fortunately most biological objects have a greater refractive index
than water and so can be stably manipulated with the waist of a laser
beam.  Ashkin {\em et al.} (1987) showed that live viruses and
bacteria could be manipulated with microscope based infra-red optical
tweezers.  The list of biophysical applications to date includes:
manipulation of single DNA molecules (Chu 1991) and of human gametes
(Tadir {\em et al.} 1991), cell micro-surgery (Steubing {\em et al.}
1991), and studies of motor proteins (Ashkin {\em et al.} 1990) and
mitotic chromosomes (Liang {\em et al.} 1991).

\subsection{Dissipative forces} \label{dissipative}
Quantizing the field with plane wave spatial modes $\exp(i \underline{k}
\cdot \underline{r} )$ gives field quanta, called photons, with momentum
$\hbar \underline{k}$.  The dissipative, or scattering, force is then
particularly simple to understand.  Each photon absorption gives a momentum
kick
$\hbar \underline{k}$ to the absorber.  Each subsequent spontaneous emission
gives a kick of the same magnitude but in a random direction,
producing no average force.  However the resultant momentum
fluctuations are what limit cooling, as we shall see in section
\ref{Dopplercool}.

The dissipative force from a laser can slow a counter-propagating beam
of atoms.  The changing doppler shift as the atoms slow can be
overcome by chirping the laser frequency or by Zeeman shifting the
atomic transition frequency with a spatially varying magnetic field.
A slowed beam may be transversely cooled by two-dimensional optical
molasses.  This consists of two lasers counter-propagating
transversely to the atomic beam and tuned below resonance - so called
red detuning.  Transverse motion doppler shifts the
counter-propagating beam towards resonance and the co-propagating beam
away from resonance.  Hence more counter-propagating photons are
absorbed and the atom slows.  This process is called doppler cooling.

There is a limit to the temperature achievable by
doppler cooling - the so called doppler limit
\begin{equation}
	k_{B} T_{Dop} = \frac{ \hbar \gamma }{2}
	\quad , \quad \mbox{Doppler limit},
	\label{dopplerlim}
\end{equation}
where $\gamma$ is the transition linewidth or inverse lifetime.  Note
that here and hereafter we use ``temperature'' in a conventional sense
to denote the mean energy without implying thermal equilibrium.  For
the D line of Na, $T_{Dop} \approx$ 240 $\mu$K. The doppler limit is
interesting because it depends only on the linewidth.  The molasses
cools the atoms until their kinetic energy equals half the transition
energy uncertainty $\hbar \gamma$ (Stenholm 1986).  This is reasonable
since the atoms are inelastically scattering photons.  Scattered
photons will typically differ in energy by the linewidth.  The
difference between the scattered photon energy and the absorbed photon
energy appears as the atomic kinetic energy.  In section
\ref{Dopplercool} we derive the doppler limit by balancing the cooling
rate with the heating rate due to the spontaneous emission recoils.

\subsection{Conservative forces \& semiclassical dressed states}
 \label{conservative}
Forces arise when there is a spatial gradient in the expectation value
of a system's energy, i.e.  Hamiltonian.  A two-level atom interacting
with the electric field $\underline{E}(\underline{r},t)$ has the
Hamiltonian (Allen and Eberly 1987),

\begin{equation}
	H = \hbar (\omega_{A}/2) ( |e\rangle \langle e| -
	|g\rangle \langle g| ) -\underline{d} \cdot
	\underline{E}(\underline{r},t) \quad ,
	\label{JCham}
\end{equation}
where $|g\rangle$ and $|e\rangle$ are the ground and excited atomic
states respectively, $\hbar \omega_{A}$ is the transition energy, and
$\underline{d}$ its electric dipole moment operator.  Assuming no spatial
dependence of $\omega_{A}$ the Heisenberg equation of motion for the
momentum $\underline{p}$ is
\begin{equation}
\underline{F}=
	\frac{d \underline{p}}{dt} = \frac{1}{i \hbar} [\underline{p},H] =
	-\underline{\nabla} H = \nabla( \underline{d}
	\cdot \underline{E}(\underline{r},t) ) \quad ,
	\label{heisenbergmom}
\end{equation}
since the time derivative of the momentum is the force, $\underline{F}$.
Forces arising from a gradient in the Hamiltonian are conservative,
there is no energy dissipation as in doppler cooling.  The force
described by equation (\ref{heisenbergmom}) is known as the gradient force or
dipole force.

The eigenstates of the Hamiltonian, equation (\ref{JCham}), are called
dressed states (Cohen-Tannoudji {\em et al.} 1992).  The energy
expectation value is the sum over the dressed states of the
product of their energies and populations.  Dipole forces arise
due to gradients in the dressed state energies, or due to dressed
state population gradients, or both.

In this paper we only consider the semiclassical case in which
$\underline{E}(\underline{r},t) = \underline{\cal E}(\underline{r})
\exp(-i \omega t) + \underline{\cal E}^{*}(\underline{r}) \exp(i
\omega t)$ is a classical field.  Ignoring the kinetic energy and
making the rotating wave approximation, the Hamiltonian becomes
\begin{eqnarray}
  	H_{SC} &=& \hbar (\omega_{A}/2) ( |e\rangle \langle e| -
	|g\rangle \langle g| ) \nonumber \\
	&& -i d \left( |e\rangle \langle g| {\cal E}(\underline{r})
	e^{-i \omega_{L}t} -|g\rangle \langle e| {\cal
	E}^{*}(\underline{r}) e^{i \omega_{L}t} \right) ,
	\label{hamsc}
\end{eqnarray}
where $\omega_{L}$ is the field frequency and $d$ is the dipole matrix
element related to the transition linewidth by $\gamma = 4
\omega_{A}^{3} d^{2} / 3 \hbar c^{3}$.  We eliminate the time
dependence of the field by working in the interaction picture with
$H_{0}= \hbar (\omega_{L}/2) (|e\rangle \langle e| -|g\rangle \langle
g|)$ and with interaction Hamiltonian
\begin{eqnarray}
	H_{SC,I} &=& -\hbar (\Delta /2) ( |e\rangle \langle e| -
	|g\rangle \langle g| )\nonumber \\
	&& -i d \left( |e\rangle \langle g| {\cal E} (\underline{r})
	-|g\rangle \langle e| {\cal E}^{*}(\underline{r}) \right) \quad ,
	\label{hamsci}
\end{eqnarray}
where we have defined the field-atom detuning $\Delta \equiv
\omega_{L}-\omega_{A}$.  The semiclassical dressed states are the
eigenstates of this Hamiltonian. They and their energies are
\begin{eqnarray}
	 |1\rangle \equiv
	\cos(\theta)|g\rangle +i\sin(\theta)|e\rangle  &,&
	E_{1} = -\frac{\hbar}{2} \cal{R} \quad ; \\
	 |2\rangle \equiv
	\sin(\theta)|g\rangle -i\cos(\theta)|e\rangle &,&
	E_{2} = +\frac{\hbar}{2} \cal{R} \quad ;
	\label{scestates}
\end{eqnarray}
\begin{equation}
	\cos(2\theta) \equiv -\Delta / {\cal R}, \quad
	\sin(2\theta) \equiv \Omega / {\cal R} \quad .
\end{equation}
We have introduced the Rabi frequency and the generalised Rabi
frequency
\begin{eqnarray}
	\Omega &\equiv& 2d |{\cal E}| / \hbar ,\quad \quad \quad
	\mbox{Rabi frequency}, \label{rabifreq} \\
	\cal{R} &\equiv& \sqrt{\Delta^{2} + \Omega^{2} } ,\quad
	\mbox{generalised Rabi freq.}
	\label{genrabifreq}
\end{eqnarray}
The difference between the generalised Rabi frequency and the detuning
is referred to as the light shift. It is the shift
in the transition frequency due to interaction with the light field.

Consider a ground state atom slowly moving into a laser beam.  The
adiabatic theorem (Messiah 1966) implies that it remains in the Hamiltonian
eigenstate continuously connected to the zero field ground state.
Which dressed state this is depends on the field detuning
 \begin{equation}
 	\Delta >0 \Rightarrow \cos 2\theta = -1
 	\Rightarrow \cos \theta = 0
 	\Rightarrow |2\rangle = |g\rangle  \quad ;
  \end{equation}
 \begin{equation}
 	 \Delta <0 \Rightarrow \cos 2\theta = 1
 	\Rightarrow \cos \theta = 1
 	\Rightarrow |1\rangle = |g\rangle  \quad .
 	\label{dressedequiv}
 \end{equation}
Consequently the Hamiltonian expectation value is just the
eigenvalue $ \pm \hbar {\cal R}/2$.  In the limit of detuning $\Delta$
large compared to the Rabi frequency $\Omega$
\begin{equation}
	\langle H \rangle \approx \pm \frac{\hbar}{2} {\cal R} \approx
	\pm \frac{\hbar}{2} | \Delta | \left( 1
	+ \frac{1}{2} \left( \frac{\Omega}{\Delta} \right)^{2} \right) \quad .
	\label{expH}
\end{equation}
For constant detuning only the second term contributes to
the gradient corresponding to an effective potential
\begin{equation}
	V(\underline{r}) = \frac{1}{4} \frac{\hbar \Omega^{2}}{\Delta} \quad .
	\label{fardetpot}
\end{equation}
This is the mechanical potential produced by intensity gradients of a
far detuned field.  It has the same sign as $\Delta=\omega_{L}
-\omega_{A}$.  So for $\Delta > 0$, blue detuning, it rises in high
intensity regions from which atoms are therefore repelled.  Otherwise,
for $\Delta < 0$, red detuning, atoms are attracted to high intensity
regions.

This far detuned potential has a simple interpretation in terms
of the bare atomic states.  The excited state population $P_{e}
\approx s/2$ is given by equation (\ref{expop}), and the atom is excited by
absorbing photons with energy $\hbar \Delta$ greater than that of the
excited atom.  The energy excess $P_{e} \hbar \Delta$ is exactly the
potential, equation (\ref{fardetpot}); the energy excess appears as
mechanical potential energy.

The red detuned dipole force can be used to construct an atom trap
called the far off resonance trap or FORT.  It is
essentially optical tweezers for atoms.  A FORT trap has been
demonstrated to operate at 65 nm detuning, confining Rb atoms at 0.4 mK
(Miller {\em et al.}  1993).

Since it is non-dissipative the dipole force is most suitable for
constructing atom optics devices which preserve quantum coherence.
This is a requirement for atom interferometry.
Proposed atom waveguides constructed from hollow optical fibres
utilize the dipole force to repel the atoms from the fibre walls
(Marksteiner {\em et al.} 1994).

\subsection{A sense of scale}
Let's get an idea of the scale of the
forces, masses, accelerations and so on that are typical of atom
optics.  For concreteness we consider a sodium atom which has
a mass of $M=23$ au = $ 3.8 \times
10^{-26}$ kg.  The yellow D-line transition is 3S$_{1/2}
\leftrightarrow $ 3P$_{3/2}$ with wavelength 589.0 nm, and lifetime
16.4 ns (Adams {\em et al.} 1994a).
A saturated sodium atom has 50\% probability to be excited,
so on average it absorbs, and emits, one photon
every 33 ns or $3 \times 10^{7}$ photons/s.  The impulse from the absorbed
photon momenta is $I = (3 \times 10^{7}) \hbar k = 3.4 \times 10^{-20}$
(kg m/s)/s.  The corresponding acceleration of the atom is $I/M
\approx 10^{6}$ m/s$^{2} \approx 10^{5}$ g.  At this acceleration a
500 m/s atom is brought to rest in half a millisecond over about 13 cm.

The doppler limit temperature, equation (\ref{dopplerlim}), of 240 $\mu$K
corresponds to a thermal velocity
\begin{equation}
	v_{d}=\sqrt{ \frac{\hbar \gamma}{4M} } = 30 \:\mbox{cm/s} \quad .
	\label{vdop}
\end{equation}
By comparison at a temperature of 300 K the average thermal speed in a
Maxwellian distribution is $\sqrt{8k_{B}T/\pi M} \approx 530$ m/s.

The recoil speed from a photon emission or absorption is
\begin{equation}
	v_{r} = \frac{ \hbar k}{M} = 3 \:\mbox{cm/s} \quad .
	\label{recoil}
\end{equation}
A thermal speed of 3 cm/s corresponds to a temperature of about 1
$\mu$K. This recoil limit speed will turn out to be the lower limit to
speeds attainable by cooling mechanisms for which the cold atoms
interact with light.  Since $p= h/ \lambda$ for both matter and light
the corresponding atomic de Broglie wavelength is, by momentum
conservation, just that of the photon $\lambda_{dB} = \lambda_{light}
=$ 589 nm.  The corresponding kinetic energy $Mv_{r}^{2}/2 = 1.7
\times 10^{-29}$ J. This is nearly 10 orders of magnitude smaller than
the energy of a photon $\hbar \omega = 3.4 \times 10^{-19}$ J. It
corresponds to a detuning of about $\Delta = 1.7 \times 10^{-29} /
\hbar = 1.6 \times 10^{5}$ rad/s.

\section{Cooling} \label{sec2}

A great variety of cooling methods have been demonstrated (Metcalf and
van der Straten 1994).  They can be classified with respect to two
limits representing the lowest temperatures attainable with the
particular method: the doppler limit and the recoil limit:
\begin{equation}
 T_{Dop}= \frac{\hbar \gamma}{ 2 k_{B} } , \quad \mbox{Doppler limit},
 \quad \quad T_{Recoil}= \frac{(\hbar k)^{2}}{ 2Mk_{B} } , \quad
 \mbox{Recoil limit}.
	\label{limits}
\end{equation}
For Na these are respectively 240 $\mu$K and 1.2 $\mu$K. Cooling using
doppler tuning of an atomic transition is doppler limited due to
heating by the scattered photon recoils.  Sub-doppler cooling utilizes
the multi-level nature of atoms to approach the recoil limit.  It is
limited by the fact that the coldest atoms have scattered a last
photon and hence have at least the photon recoil energy.  Sub-recoil
cooling requires passive velocity groups for the atoms to accumulate
in without scattering photons.

\subsection{Doppler cooling} \label{Dopplercool}
Doppler cooling was introduced in section \ref{dissipative}. In this
section the doppler cooling limit is derived by
balancing the viscous cooling with the recoil heating.
We assume a two-level atom with velocity $\underline{v}$ and transition
frequency $\omega_{A}$ interacting with 1D optical molasses at
frequency $\omega_{L}$.  The optical molasses consists of two
counterpropagating laser beams: beam 1 with wavevector $\underline{k}$ and
beam 2 with wavevector $-\underline{k}$.  The field-atom detuning is denoted
by $\Delta =\omega_{L} -\omega_{A}$.  From the semiclassical
Bloch equations (Allen and Eberly 1987) the steady state excited state
population is
\begin{equation}
	P_{e} = \frac{1}{2} \frac{s}{1+s} ,\quad \quad
	s \equiv \frac{ \Omega^{2} /2 }
	{ (\Delta-\underline{k}.\underline{v} )^{2} +\gamma^{2}/4 } \quad ,
	\label{expop}
\end{equation}
where $s$ is the saturation parameter, expressed in terms of
the Rabi frequency $\Omega$, equation (\ref{rabifreq}), for the atom
field interaction.

We now assume that the transition is sufficiently far from saturation
$s \ll 1$ that the absorption from each beam is independent.  Then the
force on the atom is the sum of the momentum absorption rate from each
beam
\begin{eqnarray}
	F &\approx& \hbar k \gamma P_{e,1} -\hbar k \gamma P_{e,2}
	  \nonumber \\
	 &\approx& \hbar k \gamma \frac{1}{2} ( s_{1} -s_{2} ),
	 \quad s_{i} \ll 1 \quad .
	\label{force1}
\end{eqnarray}
For the small velocities at the doppler cooling limit we can expand
$s$ in $v = | \underline{k}.\underline{v} | / | \underline{k} |$ to get
\begin{equation}
	s_{1/2} \approx s_{0} \left( 1 \pm
	\frac{2 \Delta k v}
	{\Delta^{2} +\gamma^{2}/4 } \right) \quad ,
	\label{satexp}
\end{equation}
where $s_{0}$ is the zero velocity saturation parameter.
Substituting these expansions into the force equation (\ref{force1}) gives
\begin{equation}
	F \approx \hbar k \gamma s_{0}
	\frac{2 \Delta  k} {\Delta^{2} +\gamma^{2}/4 } \: v \quad .
	\label{force2}
\end{equation}
For red detuning, $\Delta < 0$, this is a viscous type damping force
proportional to the atomic velocity.  The associated
kinetic energy cooling rate is
\begin{equation}
	\frac{d \langle E_{cool} \rangle}{dt} =
	\langle Fv \rangle = \hbar k \gamma s_{0}
	\frac{2 \Delta  k} {\Delta^{2} +\gamma^{2}/4 }
	\: \langle v^{2} \rangle \quad ,
	\label{coolrate}
\end{equation}
where the angle brackets denote an average over the atomic ensemble.
We next consider the photon recoil heating of the atom.

Each spontaneous emission kicks the atom in a random direction with
momentum magnitude $\hbar k$, producing a random walk in momentum
space.  So after $N$ emissions the mean square momentum magnitude is
$\langle p^{2} \rangle =N (\hbar k)^{2}$.  The total momentum
diffusion rate is four times this.  One factor of two comes from
spontaneous emission from the two beams.  The other factor of
two comes from the absorption, which is also a random process
(Cohen-Tannoudji 1992).  Using
the spontaneous emission rate $\gamma s_{0}/2$ we then have the
following expression for the kinetic energy heating rate
\begin{equation}
	\frac{d \langle E_{heat} \rangle}{dt} =
	4 \frac{d ( \langle  p^{2} \rangle /2M )}{dt}
	\approx 4 \frac{1}{2M} \gamma \frac{s_{0}}{2}
	(\hbar k)^{2} , \quad s_{0} \ll 1,
	\label{heatrate}
\end{equation}
where we have used the zero velocity saturation parameter, as our goal
is to find the temperature minimum.  At equilibrium the sum of this
heating rate and the cooling rate, equation (\ref{coolrate}), is zero
\begin{equation}
	\frac{d \langle E_{cool} \rangle}{dt}
	+\frac{d \langle E_{heat} \rangle}{dt} = 0
\quad \Rightarrow \quad
	\frac{2 \Delta}{\Delta^{2} +\gamma^{2}/4 }
	\langle v^{2} \rangle +\frac{\hbar}{M} =0 \Rightarrow
	\nonumber
\end{equation}
\begin{equation}
	 \langle \frac{1}{2} M v^{2} \rangle =
	-\frac{\hbar}{2} \frac{\Delta^{2} +\gamma^{2}/4 }{2\Delta}
	= -\frac{\hbar \gamma}{8} \left( \frac{2 \Delta}{\gamma}
	+\frac{\gamma}{2 \Delta} \right)
	\label{meanke}
\end{equation}
The last bracketed expression has a minimum of $-2$ at $2 \Delta /
\gamma = -1 \Rightarrow \Delta = -2\gamma$, so this detuning gives the
minimum mean kinetic energy.  This defines the
doppler limit energy $k_{B} T_{Dop}/2$, so
\begin{equation}
	T_{Dop} = \frac{\hbar \gamma}{2 k_{B}} \quad .
	\label{dopplerlim2}
\end{equation}
A result explained heuristically at the end of section
\ref{dissipative}.

\subsection{Sisyphus cooling}
The doppler cooling discussed in the previous section only works for
light intensities below saturation. Above saturation stimulated
processes become important and cooling only occurs for blue detuning !
This blue cooling is stronger than doppler cooling but does
not achieve as low temperatures (Aspect {\em et al.} 1986). This is because
although the damping is stronger so are the fluctuations.

It is an example of a more general mechanism called Sisyphus cooling
after the mythological Sisyphus, king of Corinth, who was doomed by
Zeus to roll a rock uphill forever.  In a standing wave the Rabi
frequency $\Omega$ varies with the spatial mode function from zero at
the nodes (zero field) to a maximum at the anti-nodes.  Hence so does
the energy splitting of the semiclassical dressed states $\hbar
\cal{R}$, equation (\ref{genrabifreq}).  Cooling occurs because, as we
shall see, spontaneous emission is most likely to occur from energy
maxima into energy minima.  The atom then moves up the energy hill and
is again most likely to emit at the maximum.  On average kinetic
energy is converted to atom-field interaction energy which is lost by
spontaneous emission.

A detailed explanation using dressed atomic states with a quantized
field has been given by Dalibard and Cohen-Tannoudji (1985).
By contrast we use semiclassical dressed states, which were introduced
in section   \ref{conservative}.  Any
dressed state with an excited state component can spontaneously emit
into any dressed state with a ground state component.  In general a
dressed state can spontaneously emit into itself as well as into the
other dressed state.

We now explain blue Sisyphus cooling with reference to Fig.
\ref{fig2}.  In a standing wave field the electric field is zero at
the nodes and is maximum at the anti-nodes.  So the generalised Rabi
frequency, equation (\ref{genrabifreq}), varies from $| \Delta |$ at
the nodes to $\sqrt{\Delta^{2}+\Omega^{2}}$ at the anti-nodes.
$|2\rangle$ is the higher energy state and for blue detuning, $\Delta
> 0$, it equals the ground state $|g\rangle$ at a node.  It increases
in energy and mixes in the excited state $|e\rangle$ as the field
increases.  $|1\rangle$ is the lower energy state and for blue
detuning it equals $|e\rangle$ at the nodes and it decreases in energy
and mixes in the ground state with increasing field.  Consequently at
anti-nodes $|2\rangle$ has its maximum energy and $|1\rangle$ its
minimum energy while at nodes the opposite holds.  But $|2\rangle$ is
most likely to spontaneously emit at anti-nodes since there it has the
largest excited state component.  It can decay either to itself, in
which case only the photon energy $\hbar \omega_{L}$ is lost, or to
$|1\rangle$, in which case the larger energy $\hbar (\omega_{L}
+\cal{R})$ is lost.  (Recall that the dressed states are defined in a
picture rotating with frequency $\omega_{L}$, so $\hbar \omega_{L}$ is
the energy zero).  In contrast the dressed state $|1\rangle$ is most
likely to spontaneously emit at nodes, where it is exactly the bare
excited state, $|e\rangle$.  At a node it has no ground state
component and hence can only decay to $|2\rangle$, losing energy
$\hbar (\omega_{L} -\Delta)$.  Since this is smaller than $\hbar
(\omega_{L} +\cal{R})$ more energy than $\hbar \omega_{L}$ is lost on
average per spontaneous emission.  This energy cannot be provided by
the photons absorbed from the field, which have energy $\hbar
\omega_{L}$.  It comes from kinetic energy and hence cools the atom.

The transfer of kinetic energy to atom-field interaction energy is the
Sisyphus mechanism.  An atom which has just spontaneously emitted
is most likely to be in the energy minimum for its dressed state. Any
motion increases the field-atom interaction energy until it is
most likely to spontaneously emit again at the top of the energy hill.
Thus kinetic energy is transformed into interaction energy which is
dissipated by spontaneous emission.

Consider a sequence of spontaneous emissions which start and end in
the same dressed state while passing through the other dressed state.
The most likely starting point is at an energy maximum and the most
likely end point an energy minimum.  Hence the total process of (at
least) two spontaneous emissions dissipates the energy difference
between the maximum and minimum, which is just the light shift.
Sequences which increase the energy are possible, but less likely.

The limiting temperature is determined by the heating due to
spontaneous emissions. Emission from $|2 \rangle$ is most likely to
leave the atom in $|2 \rangle$ with no cooling effect, but recoil
heating still occurs.

\subsection{Spatially dependent optical pumping}
To cool below the doppler limit, equation (\ref{dopplerlim2}), requires
multi-level atoms. Since atoms are multi-level and since
theorists like two-level atoms it is perhaps not surprising that sub-doppler
cooling was discovered experimentally (Lett {\em et al.} 1988). A common
feature of the various sub-doppler, but super-recoil, cooling schemes
is optical pumping.

Optical pumping is the transfer of population between
atomic magnetic sublevels. For definiteness we consider a total
angular momentum $J=1/2 \leftrightarrow J=3/2$ transition, Fig.
\ref{fig3}.
Circularly polarized light is either $\sigma^{+}$ or $\sigma^{-}$
with photon angular momentum $+\hbar$ and $-\hbar$
respectively.  An atom in the $m= -1/2$ ground state sub-level can be
excited into the $m= +1/2$ excited state sub-level by absorbing a
$\sigma^{+}$ photon, and into the $m= -3/2$ excited state
sub-level by absorbing a $\sigma^{-}$ photon.  Similarly the $m=
+1/2$ ground sub-level is excited into either the $m= +3/2$ or $m=
-1/2$ excited sub-levels by absorbing a $\sigma^{+}$ or $\sigma^{-}$
photon respectively. Absorption of
$\sigma^{+}$ polarized light increases the magnetic quantum number
$m$ while $\sigma^{-}$ polarized light decreases it.  The net result
of cycles of absorption and spontaneous emission in circularly
polarized light is optical pumping: the preferential population of ground
sub-levels with extremal $m$.  In this case either the $m= 1/2$ or $m=
-1/2$ sub-level.  In the following we are only concerned with weak
fields so that excited state population is negligible.

Spatially dependent optical pumping occurs when the the pumped
sub-level depends on position.  In 3D this is unavoidable and is
usually achieved with three pairs of counter-propagating laser beams.
Because the atoms are moving the state of the atom lags behind the
local steady state.  Cooling is achieved when this lag is used to
remove energy from the atoms as they are optically pumped towards
local equilibrium.  Since polarization gradient cooling is one of the
more effective of such mechanisms, we consider it next.

\subsection{Polarization gradient cooling}
Polarization gradient cooling uses counter-propagating beams with
perpendicular linear polarizations to achieve spatially dependent
optical pumping, Fig. \ref{fig4}.  This is known as the lin
perp lin configuration.  Atoms lose kinetic energy by a Sisyphus
mechanism in which the hills are due to light shifts.  3D lin
$\bot$ lin cooling to within a few recoil velocities, about 25 $\mu$K
for Na, has been achieved (Cohen-Tannoudji and Phillips 1990).

The polarization varies with a period of $\lambda /2$ from linear to
$\sigma^{+}$, to the orthogonal linear, to $\sigma^{-}$, and back to
linear, Fig.  \ref{fig4}.  The linear polarization is at angle $\pi
/4$ to the input linear polarizations.  This can be seen from the
expression for the electric field in the lin $\bot$ lin configuration
\begin{equation}
	\underline{E} = \underline{x} \sin (kz -\omega t)
	+ \underline{y}\sin (kz +\omega t),
	\label{linperplin}
\end{equation}
where $\underline{x}$ and $\underline{y}$ are unit vectors in the $x$ and $y$
directions. Using $k=2\pi/\lambda$ and some trigonometric
identities gives
\begin{equation}
	\underline{E} = (\underline{y}-\underline{x}) \cos (2\pi
	z/\lambda) \sin(\omega t) + (\underline{x}+\underline{y}) \sin
	(2\pi z/\lambda) \cos(\omega t) \quad .
	\label{linperplin2}
\end{equation}
At $z=0$ the field is linearly polarized in the
$\underline{y}-\underline{x}$ direction.  At $z=\lambda /8$,
$\sin(2\pi z/\lambda) = \cos(2\pi z/\lambda) = 1/\sqrt{2}$ and the
magnitude of the field is constant in time and hence it is circularly
polarized.  At $z=\lambda /4$, $\sin(2\pi z/\lambda) = 1$ and the
field is linearly polarized in the $\underline{x}+\underline{y}$
direction.  At $z=3 \lambda /8$, $\sin(2\pi z/\lambda) = 1/\sqrt{2}$
and $\cos(2\pi z/\lambda) = -1/\sqrt{2}$,  and the field is again
circularly polarized, but this time with the opposite rotation sense.
This polarization gradient produces spatially dependent
optical pumping.

Lin $\bot$ lin cooling uses light shifting of the ground sub-levels.
Light shifting was discussed for two-level transitions in section\
\ref{conservative}.  The total shift of the $m= \pm 1/2$ sub-levels is
the sum of the shifts due to each of the two transitions to which they
are coupled by the lasers. The proportion of $\sigma^{-}$ and
$\sigma^{+}$ varies spatially.  And since each transition has a
different coupling strength, as determined by the Clebsch-Gordon
coefficients, the light shifts of the ground sub-levels depend on
position.

These shifts are negative for red detuning and for the $m= -1/2$
sub-level are maximum in magnitude where the field is $\sigma^{-}$
polarized, and minimum where it is $\sigma^{+}$.  The opposite is true
for the $m= +1/2$ sub-level, Fig.  \ref{fig4}.  At the minimum light
shift positions optical pumping into the other sub-level is strongest.
The result is that atoms move uphill until they are optically pumped
into the other sub-level.

This pumping takes energy out of the atom provided the atomic state
lags the local equilibrium by somewhat less than the optical pumping
time $\tau_{p}$.  This is the average time for optical pumping to
transfer population between the ground sub-levels and is proportional
to the intensity. Hence lin $\bot$ lin cooling works best for velocities
$v_{ideal}$ such that the atom travels $\lambda /4$ in $\tau_{p}$,
$v_{ideal} \approx \lambda /(4 \tau_{p})$. Atoms which are moving too fast
do not respond to the local pumping and hence do not lose energy.
Atoms which are moving too slowly are pumped to the other sub-level
before reaching the top of the hill and hence lose less energy.

Polarization gradient cooling does not achieve the recoil limit.
Cooling closer to the recoil limit can be achieved after trapping the
atoms in an optical lattice.  Adiabatically expanding the lattice
cools the atoms.  Reduction of Cs temperature from the 3 $\mu$K
achievable with polarization gradient cooling to 0.7 $\mu$K has been
demonstrated with adiabatic cooling (Kastberg {\em et al.} 1995).
This is short of the Cs recoil limit of 0.13 $\mu$K.

\subsection{Velocity selective coherent population trapping - VSCPT}
For cooling below the recoil limit the problem is to avoid the recoil
from the last photon emitted from determining the temperature.
There are two successful approaches: momentum diffusion into a passive
zero velocity group, and evaporation of the hottest atoms, avoiding
photons altogether.  Velocity selective coherent population trapping
(VSCPT) and Raman cooling use the first approach.

In this section we discuss a simple model of 1D VSCPT based on a
three-level lambda transition.
More detailed discussions and the extension to 3D
can be found in the 1990 Les Houches lectures of Cohen-Tannoudji
(1992).  2D VSCPT cooling of metastable He to
250 nK, 16 times below the recoil limit, was reported in 1994
(Lawall 1994).

We assume that the ground levels are degenerate and that the two
counter-propagating lasers have the same frequency, and opposite
circular polarizations, Fig.  \ref{fig5}.  The interaction picture
Hamiltonian is
\begin{equation} \label{VSCPTham}
    H_{VSCPT} = \frac{p^2}{2M}
    -\hbar \Delta |3 \rangle \langle 3 |
    +d {\cal E} \left[e^{ik z}  |3\rangle
    \langle 1 | + e^{-ik z} |3\rangle
    \langle 2 |+\mbox{h.c.}\right] \quad ,
\end{equation}
where for simplicity we have assumed that the two transitions have
the same dipole moment, $d$.  Note that the plane wave spatial
dependence of the mode in the direction of propagation $z$ has been
made explicit.  A set of states closed under this Hamiltonian is $\{
|\mbox{NC},p \rangle, |\mbox{C},p \rangle, |\mbox{3},p \rangle \}$. The
first two states are defined by
\begin{eqnarray}
	|\mbox{NC},p \rangle \equiv \frac{1}{\sqrt{2}} (
	|1,p-\hbar k \rangle -|2,p+\hbar k \rangle ) \quad  ,\\
	|\mbox{C},p \rangle \equiv \frac{1}{\sqrt{2}} (
	|1,p-\hbar k \rangle +|2,p+\hbar k \rangle )  \quad .
	\label{VSCPTclosed}
\end{eqnarray}
The ``NC'' stands for Non-Coupling and the ``C'' for
Coupling.  The interesting thing is the action of the
Hamiltonian equation (\ref{VSCPTham}) on $ |\mbox{NC},p \rangle$.  Since
$\exp(ikz)$ generates a momentum space displacement of the momentum
eigenstate $|p\rangle$ to $|p+\hbar k\rangle$,
\begin{eqnarray}
	H_{VSCPT} |\mbox{NC},p \rangle &=&
	 \frac{1}{\sqrt{2}} \left\{
	 \frac{(p-\hbar k)^{2}}{2M} |1,p-\hbar k \rangle
	  - \frac{(p+\hbar k)^{2}}{2M} |2,p+\hbar k \rangle
	 \right\} \nonumber \\
	 &=& \left\{ \frac{p^{2}}{2M} + \frac{ (\hbar k)^{2}}{2M} \right\}
	|\mbox{NC},p \rangle
	 -\frac{\hbar k p}{M} |\mbox{C},p \rangle	\quad .
	\label{VSCPTaction}
\end{eqnarray}
For $p=0$ it is an eigenstate of the kinetic energy and has no
coupling to any other state.  This is called a dark state because it
does not interact with the field.  VSCPT works by accumulating atoms
in the non-coupling states $|\mbox{NC},p \approx 0 \rangle$.  The
non-coupling states are populated via momentum diffusion due to
spontaneous emission from the excited state.  The coupling amplitude
out of them, into $|\mbox{C},p \rangle$ states is proportional to $p$
and hence their lifetime increases as their momentum $p$ decreases.

In principle there is no minimum temperature for VSCPT. The range of
populated non-coupling states near $p=0$ decreases with time as they
are pumped out via the coupling state.  This is because the momentum
diffusion into a particular momentum group is independent of its
momentum, while the $|\mbox{NC},p \rangle \leftrightarrow |\mbox{C},p
\rangle$ coupling amplitude is proportional to $p$, and hence vanishes
as $p \rightarrow 0$.

The dark state $|\mbox{NC},p=0 \rangle$ is a quantum mechanical
superposition of the atom moving to the left and to the right with the
recoil momentum $\hbar k$.  In the 2D case the corresponding dark state
is a superposition of four atomic momentum states.  These states were
observed in a Paris experiment (Lawall {\em et al.} 1994). To confirm the
superposition, rather than an incoherent mixture, would
require interferometry, which remains to be done.

\subsection{Raman cooling}
Like VSCPT Raman cooling works by isolating the atoms near zero
velocity from the cooling fields. However there is no dark state in
Raman cooling, rather the extremely narrow linewidths possible for
Raman transitions enable tuning of the Raman pulses so that they do
not affect the zero velocity group.

Because the energy difference between the levels in a Raman transition
can be very small they can have narrow linewidths.  This can be used
to select out correspondingly narrow velocity groups, and push them
around.  Kasevich and Chu (1992) used this method to push atoms
towards zero velocity and achieved 100 nK for Na in 1D, which is 10
times below recoil.  In 2D and 3D they did not quite get to recoil
although they beat polarization gradient cooling by a factor of nearly
20 (Davidson {\em et al.} 1994).

Raman cooling uses a sequence of Raman pulses from counter-propagating
lasers to push atoms towards zero velocity.  Each Raman transition
changes the atomic velocity by $2 \hbar k$.  The pulses are designed
so zero velocity atoms are unaffected, and the accuracy with which
this can be achieved is one of the limits to the cooling.  Since many
Raman cycles are needed for cooling, the atoms must be repumped to
their original state.  The zero velocity state is populated randomly
by atoms which have zero velocity after repumping.

\subsection{Evaporative cooling}
Evaporative cooling has achieved the lowest 3D temperatures to date,
170 nK for Rb (Anderson {\em et al.} 1995), and 100 nK for Li (Bradley
{\em et al.} 1995).  These temperatures should be compared to the
recoil limit temperatures, equation (\ref{limits}), of
$T_{Recoil}=230$ nK for Rb and $T_{Recoil}=5.5$ $\mu$K for Li.  This
is what made Bose-Einstein condensation of these atoms possible.  The
cooling is achieved by bleeding off the highest energy atoms from the
trap.  Thermalisation of the remaining atoms by elastic collisions
produces cooling.  After a subsequent adiabatic expansion Anderson et
al.\ (1995) reported a temperature of 20 nK.

The trap is magnetic, section \ref{magtrap}, so that no optical
heating occurs.  Only atoms in magnetic states such that
their magnetic dipole is attracted to the trap centre are trapped.  If
the magnetic state is changed the atoms are no longer trapped and they
escape.  Such spin flips are induced by RF fields tuned to a Zeeman
splitting frequency.  Since this frequency depends on the local
magnetic field strength, spin flips can be selectively induced in the
outer parts of the traps.  These are precisely the regions occupied by
the high energy atoms.  Progressive cooling is achieved by ramping
down the RF frequency.  This moves the resonance region towards the
centre of the trap, where fields are lowest, and allows colder atoms
to evaporate.
\section{Trapping} \label{sec3}

The Optical Earnshaw theorem states that atoms for which the
dissipative force, section \ref{dissipative}, is proportional to the
intensity cannot be trapped by static configurations of laser beams
(Ashkin and Gordon 1983).  It is true for the same reason as the
classical Earnshaw theorem is true, section \ref{ion}.  In a region of
space without sources or sinks for light there is as much energy
flowing in as out, and the dissipative force is in the same direction
as the energy flow.

Fortunately the conditions of the theorem can be violated by adding
external fields (Pritchard {\em et al.} 1986).  In the case of the highly
successful magneto-optical trap this is an inhomogeneous magnetic field which
 Zeeman shifts the transition so that the force is not just
proportional to the local intensity.

\subsection{Optical molasses} \label{molasses}
Optical molasses, discussed in section \ref{Dopplercool}, does not trap
atoms.  However because the atomic motion is diffusive rather
than ballistic the atoms can take seconds to travel centimeters
(Chu {\em et al.} 1985).
The dynamics is that of Einstein/Langevin Brownian
motion (Gardiner 1985). Using the damping constant of equation (\ref{force2}),
\begin{equation}
	 \alpha \equiv\hbar k \gamma s_{0}
	\frac{2 \Delta  k} {\Delta^{2} +\gamma^{2}/4 } \quad  ,
	\label{dampconst}
\end{equation}
we have the Langevin equation for the 1D position $x$ of a doppler cooled atom
\begin{equation}
	M \frac{d^{2}x}{dt^{2}} = \alpha v +N \quad  ,
	\label{Langevin}
\end{equation}
where we have introduced the zero mean noise term $N$.  An equation
for $\langle x^{2} \rangle$ may be found after multiplying equation
(\ref{Langevin}) by $x$ and using $\langle M v^{2}/2 ,
\rangle = k_{B} T/2$
\begin{equation}
	\frac{d \langle x^{2} \rangle}{dt} = \frac{2 k_{B}T}{\alpha}
	+C \exp(-\alpha t /M) \quad  ,
	\label{rmseq}
\end{equation}
where $C$ is an integration constant. For long times the exponential on
the right hand side can be ignored and then
\begin{equation}
	\langle x^{2} \rangle -\langle x_{0}^{2} \rangle =
	\frac{2 k_{B}T}{\alpha} \: t \quad  .
	\label{xrms}
\end{equation}
For Na the root-mean-square displacement over one second is about a few
centimeters for $T$ = 100 $\mu$K.

\subsection{Magneto-optical traps: MOTs}
 Magneto-optical traps, or MOTs, are now quite common (Wieman {\em et
 al.} 1995).  They are also known as Zeeman-optical traps, or ZOTs.
 They use an inhomogeneous magnetic field to produce a position
 dependent scattering force.  The magnetic field Zeeman shifts
 transitions towards or away from resonance with red-detuned,
 counter-propagating, oppositely circularly polarized laser beams.
 Consider a $J=0 \leftrightarrow J=1$ system.  The two transitions
 coupling to circularly polarized light are shifted by positive and
 negative magnetic fields as shown in Fig.\ \ref{fig6}.  The magnetic
 field generated by anti-Helmholtz coils is also shown in Fig.\
 \ref{fig6}.  For an atom displaced towards the $\sigma^{-}$ polarized
 beam the magnetic field becomes more positive and it Zeeman shifts
 the $\sigma^{-}$ transition into resonance and the $\sigma^{+}$
 transition away from resonance.  Hence the scattering from the
 $\sigma^{-}$ beam dominates and the atom is pushed back towards the
 centre of the trap.  The magnetic field and laser polarizations are
 such that atoms are pushed back to the centre after displacement in
 any direction.

The MOT lasers cool the atoms too.  An atom stopping at its turning
point is accelerated back towards the centre.  However once the
doppler shift $kv$ exceeds the linewidth $\gamma$ it is no longer
accelerated.  So the maximum rebound speed is about $\gamma /k$,
corresponding to a temperature of about $M (\gamma /k)^{2}/k_{B}$
(Metcalf and van der Straten 1994).
This is much higher than the doppler limit, equation (\ref{dopplerlim2}).

\subsection{Ion traps} \label{ion}
Ion traps have certain advantages over optical traps (Blatt 1992).
Most importantly the trapping does not rely on interaction with light.
The potentially extremely small dissipation in ion traps has allowed
the demonstration of quantum logic gates (Monroe {\em et al.} 1995).
They are one of the more promising ways of realizing quantum computers
(Cirac and Zoller 1995).

Since the divergence of the electric field in free space is zero any
region must have field lines both going in and coming out.
Equivalently there can be no maxima or minima of the electrostatic
potential in free space.  This is the content of Earnshaw's electrostatic
theorem. Consequently an ion cannot be trapped by
electrostatic fields alone. This problem has been solved in two ways:
by adding a magnetic field, the Penning trap, or by
oscillating the electric field, the Paul trap.
A quadrupole potential
\begin{equation}
	V = V_{0} ( r^{2} -2z^{2}) \quad  ,
	\label{quadpot}
\end{equation}
where $r$ is the radial coordinate, is produced by the ring and endcap
electrode configuration shown in Fig. \ref{fig7}.

The Penning trap achieves radial confinement with an axial magnetic
field, in the $z$ direction.  This inhibits radial motion by
converting it to circular cyclotron motion about the magnetic field
lines.  The overall motion of the ion is the combination of a slow
magnetron rotation ( $\approx$ 10 kHz) about the $z$ axis and the fast
cyclotron motion ( $\approx$ 1000 kHz) about the local magnetic field
lines.  The electrostatic field provides confinement in the $z$
direction. The axial oscillation frequency lies between these two.

The Paul trap adds RF time dependence to the potential equation
(\ref{quadpot}) of the form $V_{0}=V_{DC}+V_{AC} \cos \omega t$.  This
allows a dynamic stability.  Again the ion motion has two components;
slow approximately harmonic motion in the axial and radial directions,
and rapid micromotion oscillation about the local position at the
driving frequency ( $\approx$ 10 MHz).

\subsection{Magnetic traps} \label{magtrap}
Magnetic traps confine atoms by exerting a force on the atomic
magnetic dipole moment.  Many different trap configurations have been
proposed (Bergeman {\em et al.} 1987).  Since they avoid laser
heating, and the Coulomb repulsion of ions, magnetic traps have been
used to Bose-Einstein condense atoms.

The force is the gradient of the
dipole energy $-\underline{\mu} \cdot \underline{B}$, where
$\underline{\mu}$ is the dipole moment of the magnetic sub-level of
the atom and $\underline{B}$ is the magnetic field.  For sufficiently
slow atomic motion the the magnetic sub-level adiabatically follows
the changing direction of the magnetic field as it moves around the
trap.  For example an atom in the local ground state remains in the
local ground state.  Consequently its magnetic moment changes.
Atoms in magnetic sub-levels whose Zeeman energy $-\underline{\mu}
\cdot \underline{B}$ decreases with magnetic field strength are forced
towards the field minimum.

Anderson {\em et al.}\ (1995) used a variation on the simple
quadrupole trap formed with anti-Helmholtz coils (Midgall {\em et al.}
1985), called the time-averaged orbiting potential, or TOP, trap.
This overcomes a problem with the simple quadrupole trap caused by its
minimum being zero magnetic field.  Close to $B=0$ the Zeeman levels
are nearly degenerate and the atomic state may no longer adiabatically
follow the magnetic field. Non-adiabatic transitions to non-trapped magnetic
sublevels can occur, leading to trap loss.

This problem is particularly bad in quadrupole traps because the field
increases linearly in all directions away from the minimum, so the
field derivative is not zero at the field zero.  The TOP trap adds an
RF field which rotates the field zero in a circle.  The time averaged
field then has a non-zero, quadratic minimum where the zero previously
was.  Atoms are still lost from the instantaneous zero, but they are
high energy atoms with respect to the time-averaged potential and
hence contribute to evaporative cooling.

Bradley {\em et al.}\ (1995) used an arrangement of six permanent
magnets to produce a field with a non-zero minimum in which they
reported Bose-Einstein condensation of Li.

\subsection{Gravitational traps}
Gravity is an important consideration for cold atoms.  The energy
gained by a Na atom falling through 1 mm is about $4 \times 10^{-28}$
J, or more than twenty times the recoil energy.  Gravity can be used
to trap atoms in a trampoline configuration.  Ten bounces of about 3
mm height have been observed experimentally off a parabolic evanescent
wave reflector (Aminoff {\em et al.} 1993).  This corresponds to a
trapping time of about 0.1 s.  This result can probably be
substantially improved by enhancing the evanescent field.  Confinement
times at least 50 times longer have been obtained using two sheets of
blue detuned light to form a ``V'' in which atoms bounce (Davidson
{\em et al.} 1995).

Other reflector shapes such as pyramids and cones have been analyzed
(S\"{o}ding {\em et al.} 1995; Dowling and Gea-Banacloche 1995).
Recoil limited Sisyphus cooling mechanisms
which utilize gravity can be incorporated.  For example if the
sub-level changes during the bounce the average force on the way out can be
less than that on the way in (Ovchinnikov {\em et al.} 1995), leading
to an inelastic, cooling bounce.

\subsection*{Acknowledgment}
Thanks to J.\ Hope for reading the manuscript and suggesting
improvements. Thanks to the participants in the Workshop on Atom
Optics for providing corrections.

\section*{References}
\setlength{\parindent}{0pt}

Adams, C.S., Sigel, M., and Mlynek, J. (1994a).  {\it Physics
	Reports} {\bf240}, 1.

Adams, C.S., Carnal, O., and Mlynek, J.(1994b).  {\it Advances in
Atomic, Molecular, and Optical Physics} {\bf34}, 1.

Allen, L., and Eberly, J.H. (1987). ``Optical resonance and two-level atoms''
 (Dover: NY).

Allen, L., Beijersbergen, M.W., Spreeuw, R.J.C., and Woerdman J.P.
(1992).  {\it Phys.  Rev.  A} {\bf 45}, 8185.

Aminoff, C.G., Steane, A.M., Bouyer, P., Desbiolles, P., Dalibard, J.,
and Cohen-Tannoudji, C. (1993). {\it Phys.  Rev.  Lett.}
	{\bf 71}, 3083.

Anderson, M.H., Ensher, J.R., Matthews, M.R., Wieman,
	C.E., and Cornell, E.A. (1995). {\it Science} {\bf269}, 198.

Ashkin, A. (1972). {\it Scientific American}, Feb. p.63.

Ashkin, A., and Gordon J.P. (1983). {\it Opt. Lett.}  {\bf 8}, 511.

Ashkin, A., Dziedzic, J., and Yamane, T. (1987). {\it Nature}
	{\bf 330}, 769.

Ashkin, A., Schutze, K., Dziedzic, J.M., Euteneuer, U., and Schliwa,
M. (1990).  {\it Nature} {\bf 348}, 346.

Aspect, A., Dalibard, J., Heidmann, A., Salomon, C., and Cohen-Tannoudji,
C. (1986). {\it   Phys.  Rev.  Lett.}  {\bf 57}, 1688.

Bergeman, T., Erez, G., and Metcalf, H.J. (1987).  {\it
Phys.  Rev.} A {\bf 35}, 1535

Bjorkholm, J.E., Freeman, R.R., Ashkin, A., and Pearson, D.B. (1978).
{\it Phys.  Rev.  Lett.} {\bf41}, 1361.

Blatt, R. (1992).
	{\it In} '''Fundamental systems in quantum optics: Les Houches
	session LIII'', p. 253 (North-Holland: Amsterdam).

Bradley, C.C., Sackett, C.A., Tollett, J.J., and
	Hulet, R.G. (1995). {\it Phys. Rev. Lett.} {\bf75},
	1687.

Chu S., Hollberg, L., Bjorkholm, J.E., Cable, A., and Ashkin, A. (1985).
	{\it Phys.  Rev.  Lett.}  {\bf 55}, 48.

Chu, S. (1991). {\it Science} {\bf 253}, 861.

Cirac, J.I., and Zoller, P. (1995). {\it Phys. Rev. Lett.} {\bf 74}, 4091.

Cohen-Tannoudji, C., and Phillips, W.D. (1990).
{\it Physics Today} {\bf 43} No. 10, 33.

Cohen-Tannoudji, C. (1992).  ``Atomic motion in laser
	light'', p. 1 of ``Fundamental systems in quantum optics: Les Houches
	session LIII'', (North-Holland: Amsterdam).

Cohen-Tannoudji, C., Dupont-Roc, J., and Grynberg, G. (1992).
	``Atom-photon interactions'' (Wiley: NY).

Dalibard, J., and Cohen-Tannoudji, C. (1985).
{\it J. Opt. Soc. Am. B} {\bf 2}, 1707.

Davidson, N., Lee, H.-J., Kasevich, M., and Chu, S. (1994).  {\it
Phys.  Rev.  Lett.} {\bf 69}, 3158.

Davidson, N., Lee, H.J., Adams, C.S., Kasevich, M., and Chu, S.
(1995).  {\it Phys.  Rev.  Lett.} {\bf 74}, 1311.

Dowling, J.P., and Gea-Banacloche, J. (1995), ``Quantum atomic
	dots'', unpublished.

Gardiner, C. W. (1985). ``Handbook of stochastic methods'', 2nd ed.
(Springer-Verlag: Berlin).

Harris, D.J., and Savage, C.M. (1995). {\it Phys.  Rev.}  A  {\bf 51},
	3967.

He, H., Friese, M.E., Heckenberg, N.R., and
	Rubinsztein-Dunlop, H. (1995). {\it Phys. Rev. Lett.} {\bf 75}, 826.

Hope, J.J., and Savage, C.M. (1995). ``Mechanical potentials due
	to Raman transitions'', unpublished.

Kasevich, M., and Chu, S. (1992). {\it Phys.  Rev.  Lett.}
	{\bf 69}, 1741.

Kastberg, A., Phillips W.D., Rolston, S.L., and Spreeuw R.J.C. (1995).
{\it Phys.  Rev.  Lett.} {\bf 74}, 1542.

Lawall, J., Bardou, F., Saubamea B., Shimizu, K., Leduc M., Aspect,
A., and Cohen-Tannoudji, C.  (1994). {\it Phys.  Rev.  Lett.}  {\bf 73}, 1915.

Lett, P., Watts, R., Westbrook, C., Phillips, W.D., Gould, P., and
Metcalf, H. (1988).  {\it Phys.  Rev.  Lett.} {\bf61}, 169.

Liang, H., Wright, W.H., He, W., and Berns, M.W. (1991), {\it Exp.
Cell Res.} {\bf 197}, 21.

Loudon, R. (1973).  ``The Quantum Theory of Light'',1st Edition, Chap.
4 (Oxford UP: Oxford).

Mallove, E.F. and Matloff, G.L. (1989). ``The Starflight Handbook'' (Wiley:
NY).

Marksteiner, S., Savage, C. M., Zoller, P., and Rolston S. (1994).
	{\it Phys.  Rev.  A.} {\bf 50}, 2680.

Messiah, A., (1966). ``Quantum Mechanics'', volume II, chapter XVII
(North-Holland: Amsterdam).

Metcalf, H. and van der Straten, P. (1994). {\it Physics Reports}
 {\bf 244}, 203.

Midgall, A.L., Prodan, J.V., Phillips, W.D., Bergeman, T.H., and
Metcalf, H.J. (1985).  {\it Phys.  Rev.  Lett.} {\bf 54}, 2596.

Miller, J.D., Cline, R.A., and Heinzen, D.J. (1993).  {\it Phys.  Rev.
A.} {\bf 47}, 4567.

Minogin, V.G. and Letokov, V.S. (1987). ``Laser light pressure on atoms''
 (Gordon and Breach: NY).

Monroe, C., Meekhof, D.M., King, B.E., Itano, W.M., and Wineland, D.J.
 (1995), ``Demonstration of a universal quantum logic
gate'', unpublished.

Ovchinnikov, Yu.B., S\"{o}ding, C.G., Grimm, R. (1995).  {\it JETP
Lett.} {\bf 61}, 21.

Phillips, W.D., and Metcalf, H.J. (1987).  {\it Scientific American}, Mar. 36.

Pritchard, D., Raab, E.L., Bagnato, V., Wieman, C.E., and Watts, R.N.
(1986).  {\it Phys.  Rev.  Lett.} {\bf 57}, 310.

S\"{o}ding, C.G., Grimm, R., Ovchinnikov, Yu.B. (1995).  {\it Opt.
Commun.} {\bf 119}, 652.

Stenholm, S. (1986). {\it Rev. Mod. Phys.} {\bf 58}, 699.

Steubing, R.W., Cheng, S., Wright W.H., and Berns M.W.
(1991).  {\it Cytometry} {\bf 12}, 505.

Tadir, Y., Wright, W.H., Vafa, O., Liaw, L.H., Asch, R., and Berns,
M.W. (1991).  {\it Human Reprod.} {\bf 6}, 1011.

Wieman, C.E., Flowers, G., and Gilbert, S. (1995).
	{\it Am.  J. Phys.}  {\bf 63}, 317.

Wiseman, H.M, and Collett, M.J. (1995). {\it Phys. Lett. A} {\bf 202}, 246.

\newpage

\begin{figure}
	\caption{Optical ray diagrams showing the origin of dispersive forces
	on transparent macroscopic particles. The arrows represent light rays
	with thickness proportional to intensity. The columns of arrows represent
	the light beam. The left and right hand circles respectively represent
	particles with refractive index higher and lower than that of the
	surroundings. Since the light has momentum and it changes direction
	by refraction, momentum is transferred to the particles. The net
	momentum transfer is determined by the most intense rays, closest to the
	beam centre.}
	\protect\label{fig1}
\end{figure}

\begin{figure}
	\caption{Mechanism for blue Sisyphus cooling.  The top curve is
	the standing wave field intensity as a function of position.  The
	lower two curves are the energies $E_{1}$ and $E_{2}$ of the
	dressed states $|1 \rangle$ and $|2 \rangle$.  The vertical arrows
	denote spontaneous emissions, which are shown at their most probable
	locations. The diagonal arrows denote the atomic
	motion.  $\varepsilon$ is a number $\ll 1$.}
	\protect\label{fig2}
\end{figure}

\begin{figure}
	\caption{Optical pumping in a $J=1/2 \leftrightarrow J=3/2$
	transition.  States are labelled by their magnetic quantum number,
	$m$.  The thick double arrowed lines represent transitions induced
	by either $\sigma^{+}$ or $\sigma^{-}$ polarized light.  The thin
	single arrows represent spontaneous emission.}
	\protect\label{fig3}
\end{figure}

\begin{figure}
	\caption{Mechanism for polarization gradient lin $\bot$ lin
	cooling.  A $J=1/2$ ground level, as shown in Fig.  3, is assumed.
	The top atomic energy level diagrams indicate the local direction
	of optical pumping.  The curves show the energies of the dressed
	ground sublevels for red detuned light.  The pairs of vertical
	arrows show optical pumping at the positions where it is most
	probable.  The horizontal axis shows how the polarization of the
	field varies over half a wavelength due to the perpendicularly
	polarized counter-propagating lasers.}
	\protect\label{fig4}
\end{figure}

\begin{figure}
	\caption{Schematic diagram of the atomic levels, lasers, and
	transitions used in VSCPT.}
\label{fig5}
\end{figure}

\begin{figure}
	\caption{The magneto-optical trap. The left part of the diagram
	schematically shows Zeeman shifting of a $J=0
	\leftrightarrow J=1$ transition associated with positive and negative
	magnetic fields. The right part schematically shows the magnetic
	field and laser configuration in a MOT. The lasers are labelled by
	their circular polarization, either $\sigma^{+}$ or $\sigma^{-}$. The
	resultant net force on atoms at two locations are shown.}
	\protect\label{fig6}
\end{figure}

\begin{figure}
	\caption{Ion trap electrodes. Only half the ring electrode is shown.
	The $z$ direction is vertical and the radial direction is horizontal.}
	\protect\label{fig7}
\end{figure}

\end{document}